\documentclass{article}
\usepackage{spconf,amsmath,graphicx,hyperref}

\usepackage{multirow}
\usepackage{booktabs} 

\title{Pay more attention to audio: mitigating imbalance of cross-modal attention in large audio language models}

\name{
\begin{tabular}{c}
Junyu Wang$^1$, Ziyang Ma$^2$, Zhengding Luo$^3$, Tianrui Wang$^{1}$, Meng Ge$^1$,\\
\textit{Xiaobao Wang$^1$, Longbiao Wang$^{1,4,\dagger}$}
\thanks{$\dagger$ Corresponding author. } 
\end{tabular}
} 

\address{
\textsuperscript{1}Laboratory of Cognitive Computing and Application, College of Intelligence and Computing, \\ Tianjin University, Tianjin, China \\
\textsuperscript{2}School of Computer Science, Shanghai Jiao Tong University, Shanghai, China\\
\textsuperscript{3}School of Electrical \& Electronic Engineering, Nanyang Technological University, Singapore\\
\textsuperscript{4}Huiyan Technology (Tianjin) Co., Ltd, Tianjin, China\\
}

\begin{document}
\fontsize{9.5}{11.5}\selectfont
\maketitle
\begin{abstract}
Large Audio-Language Models (LALMs) often suffer from audio-textual attention imbalance, prioritizing text over acoustic information, particularly in the multi-modal fusion layers of the Transformer architecture. This bias hinders their ability to fully utilize acoustic cues, causing suboptimal performance on audio reasoning tasks. 
To mitigate this, we propose \textbf{MATA}, a novel training-free method that dynamically pushes LALMs to pay \textbf{M}ore \textbf{A}ttention \textbf{T}o \textbf{A}udio tokens within the self-attention mechanism. 
Specifically, MATA intervenes post raw attention scoring, targeting only the last token in intermediate layers without introducing additional parameters or computational overhead. Experiments on the MMAU and MMAR benchmarks confirm MATA's effectiveness, with consistent performance gains. Notably, on MMAR, MATA enables an open-source model to surpass the proprietary Gemini 2.0 Flash for the first time. Our work provides an efficient solution to mitigate attention bias and opens a new research direction for enhancing the audio-processing capabilities of multi-modal models.
\end{abstract}
\section{Introduction}
\label{intro}

In the realm of artificial intelligence, deriving logical conclusions or uncovering implicit intentions from perceived information, which is known as reasoning, is a key ability that supports advanced cognitive tasks \cite{SurveyLMRM, deepseekR1, COT1, visualCOT}. 
In the field of audio processing, audio reasoning tasks require models to deeply understand aural information, such as speaker intent, emotional tone, and subtle environmental cues \cite{Audio-reasoner, Audio-Flamingo2, Audio-cot}. For instance, in a complex soundscape, a model must be able to reason that a sudden, high-pitched scream indicates distress or danger, while the consistent hum of an engine suggests an ongoing process. Effectively performing these tasks demands a comprehensive grasp of all available audio information.

The rise of large audio-language models (LALMs), which integrate large language models (LLMs) with audio encoders \cite{LTU, GAMA, salmonn, qwen2audio}, has shown great promise in handling complex audio-textual tasks. These models leverage the robust language understanding capabilities of large language models while incorporating audio information, enabling them to perform a variety of tasks, from audio captioning to complex question answering. However, despite their impressive performance, a critical challenge remains: the attention bias between audio and text modalities. Previous research has already demonstrated similar issues in large vision-language models (LVLMs) and LLMs \cite{imagebias, PAI, hallucinationinfer, hallucinationsurvey}, where models show inconsistent attention to different input types, often prioritizing text over visual or other modalities. Yet, analogous investigations into the attention dynamics between audio and text in LALMs remain scarce. This oversight is non-trivial because if LALMs disproportionately prioritize text over audio, the task-critical acoustic information will be underutilized, hindering the model’s capacity to capture underlying intent and causing suboptimal or erroneous inferences.

\begin{figure}[t]
  \centering
  \vspace{-0.1cm}
  \includegraphics[width=0.9\linewidth]{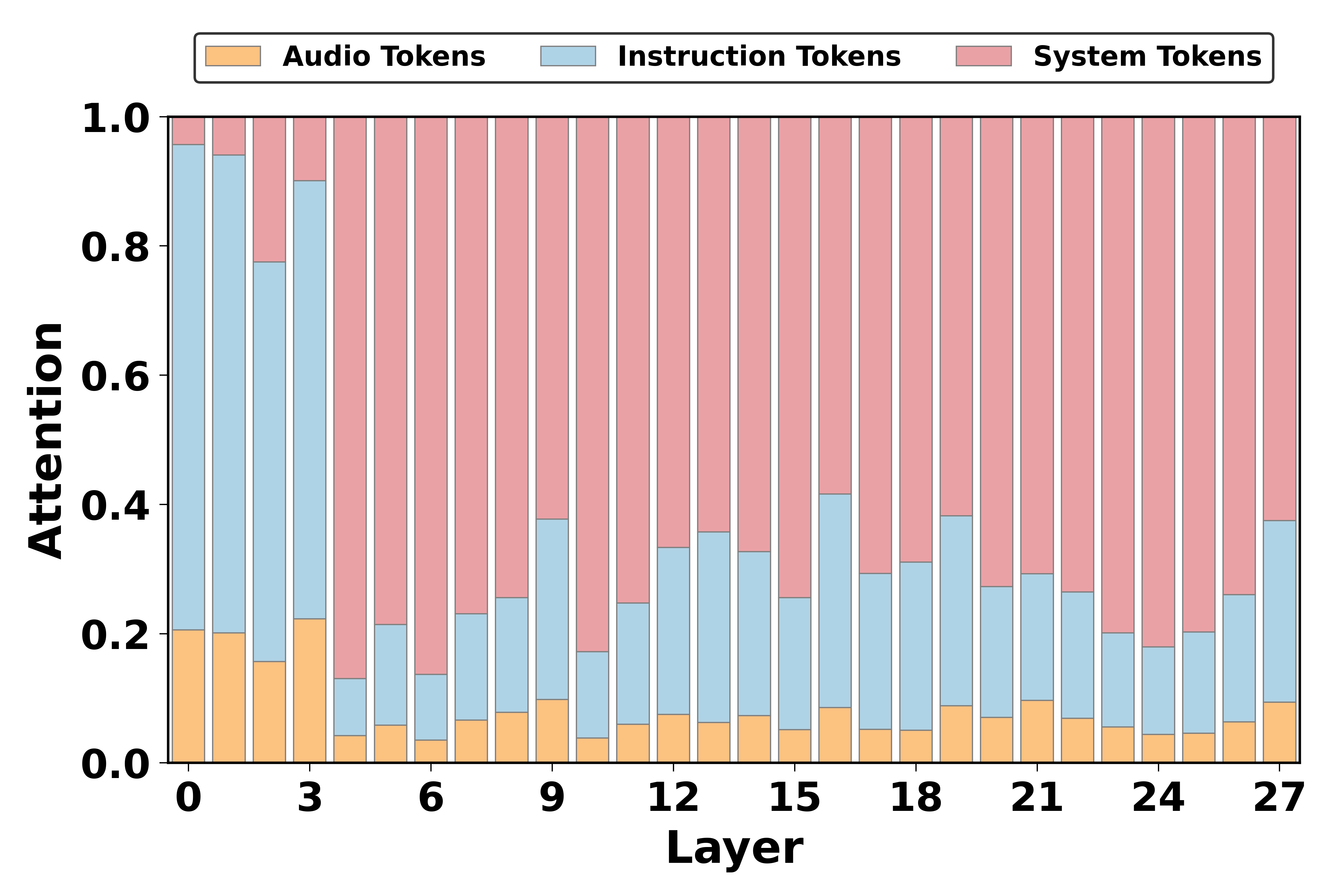}
  \vspace{-0.1cm}
  \caption{The distribution of average attention weights among all predicted tokens across different layers with Qwen2.5-Omni-7B. Instruction tokens refer to text inputs provided by users. System tokens are special model-internal tokens used to format the conversation.}
  \label{figure1}
  \vspace{-0.1cm}
\end{figure}

In this work, we are the first to systematically investigate the attention allocation of LALMs when processing multi-modal audio and text inputs. Our analysis, performed on the Qwen2.5-Omni-7B model \cite{qwen2.5omni}, reveals a significant attention disparity: the model allocates considerably more attention to text tokens than to audio tokens throughout the answer generation process. This bias is particularly pronounced in the intermediate layers of the decoder, which are responsible for the critical fusion of multi-modal information \cite{hallucinationenhance, MLLMsurvey}. As shown in Fig.~\ref{figure1}, we first save the attention distributions at each auto-regressive decoding step, then average these distributions to obtain the mean attention weight for each decoder layer. The results show that in the intermediate layers, attention to audio tokens consistently remains low, while attention to system and instruction text tokens stays high. We argue that this imbalance prevents the model from effectively utilizing the rich information embedded in the audio signals during the decoding process, which can lead to poor reasoning performance and hallucinations.

To address this challenge, we propose a novel, training-free technique to enhance the model's focus on the audio modality during inference. By directly increasing the attention weights allocated to audio tokens within the intermediate layers of the LALM's decoder, our method encourages a more audio-centric reasoning process. We evaluate our approach on two prominent audio reasoning benchmarks, MMAU \cite{mmau} and MMAR \cite{mmar}. Experimental results validate the effectiveness of our method in improving performance when integrated with open-source LALMs. Notably, when our technique is applied to the Qwen2.5-Omni-7B model fine-tuned by reinforcement learning (RL), it achieves a new state-of-the-art performance on the MMAR benchmark, surpassing even the closed-source Gemini 2.0 Flash for the first time, with an average accuracy of 66.8\%. This work highlights a fundamental limitation in current LALMs and provides a straightforward yet highly effective solution to further unleash their potential for audio reasoning tasks.

\section{RELATED WORK}
\label{related}

\subsection{Large Audio-Language Model}

The rapid evolution of LLMs has driven the development of LALMs, which extend the strengths of LLMs into the multi-modal audio-textual domain. Early works like LTU \cite{LTU} and GAMA \cite{GAMA} aligned pre-trained audio encoders with LLMs via fine-tuning, laying the groundwork for basic audio understanding. Qwen2-Audio \cite{qwen2audio} marked a significant milestone by integrating Reinforcement Learning with Human Feedback (RLHF), enhancing the model's ability to understand and respond to audio inputs in a more human-aligned manner. Building upon Qwen2-Audio, subsequent works like Audio-CoT \cite{Audio-cot} and Audio-Reasoner \cite{Audio-reasoner} introduced Chain-of-Thought (CoT) techniques to the audio-language domain. By breaking down complex audio reasoning tasks into sequential, interpretable steps, these methods partially improved the models' reasoning capabilities. More recently, omni-modal models like Qwen2.5-Omni \cite{qwen2.5omni} and Baichuan-Omni-1.5 \cite{baichuanomni} have demonstrated exceptional performance in audio-textual scenarios. Despite these notable advancements, a critical challenge remains: the inherent attention imbalance between the audio and text modalities within these models.

\subsection{Attention Weight Bias for Multi-modal Input}

Cross-modal attention bias occurs when multi-modal models disproportionately focus on one modality, like text, while neglecting others, such as audio or vision. While this bias has been extensively studied in LVLMs \cite{hallucinationbias, retrain_align}, it remains largely unexplored in LALMs. In LVLMs, this imbalance often causes hallucinations, as the models, pre-trained on vast text corpora, favor text over visual information \cite{imagebias, hallucinationsurvey}.

In contrast, the LALM domain lacks systematic research on this audio-textual attention bias. A recent study by \cite{omnir1} indirectly highlights this gap, showing that LALMs can perform competitively on audio reasoning tasks even with the audio input completely removed, relying solely on text instructions. This suggests that current LALMs have a tendency to neglect the audio modality and rely excessively on textual information, underscoring the need for deeper understanding and effective solutions.

\section{Methodology}
To address the audio-textual attention imbalance in LALMs while circumventing the high computational cost of full model retraining, we propose a approach referred to as Pay \textbf{M}ore \textbf{A}ttention \textbf{T}o \textbf{A}udio (MATA). MATA is a training-free technique that dynamically increases the model's focus on audio tokens within its self-attention mechanism. The core motivation for MATA stems from our key finding: in the intermediate layers of the Transformer architecture, the primary locus of multi-modal information integration, LALMs exhibit a significant disparity in attention allocation between text and audio tokens. By intervening at this critical stage, we can guide the model towards a more balanced multi-modal representation before the information propagates to deeper layers for output decoding.

Current LALMs commonly employ an autoregressive decoding paradigm for text generation. Within this framework, the model produces output tokens sequentially: the prediction of each new token depends on the context provided by all previously generated tokens, as well as the input multi-modal tokens. The self-attention mechanism, a core component of Transformer-based LALMs, computes raw attention weights through a dot-product operation between query \(Q\) and key \(K\) states. To prevent numerical instability from large dot-product values, these raw weights are scaled by the square root of the key dimension, with the full attention computation formulated as:
\begin{equation}
\begin{aligned}
\mathbf{A}_{(Q,K,V)} = \text{softmax}\left(\frac{Q K^{T}}{\sqrt{d_k}}\right)
\end{aligned}
\end{equation}
Subsequently, the scaled scores undergo softmax normalization to obtain normalized attention weights, which explicitly determine the contribution of each input token to the current prediction step.

Our MATA method intervenes in the self-attention computation process after the raw attention scores are calculated but before the softmax function is applied. This timing is crucial as it ensures the enhancement of audio token attention without disrupting the overall attention normalization process. Furthermore, MATA is applied only to the attention scores of the last token in the sequence. This token is a key determinant of the next output, and thus benefits most from a balanced audio-textual context. We define a small hyperparameter \(\alpha\) to control the enhancement strength. The modified attention scores, denoted as $\mathbf{\hat{A}}$, are calculated as follows:
\begin{equation}
    \mathbf{\hat{A}}_{h,i,j} = 
    \begin{cases}
        (1+\alpha) \cdot \mathbf{A}_{h,i,j}, & \text{if } i = L-1 \text{ and } j \in [a_s, a_e] \\
        \mathbf{A}_{h,i,j}, & \text{otherwise}
    \end{cases}
\end{equation}
where $h$, $i$, and $j$ denote the attention head, query position, and key position, respectively. $L$ is the total sequence length, and $[a_s, a_e]$ represents the start and end indices of the audio tokens. The modified scores are then normalized via the softmax function and used to compute the final attention output with the value \(V\) tokens. It is important to note that MATA is implemented by modifying the forward pass of the LALM's self-attention module, requiring no additional trainable parameters. This ensures that the computational overhead is negligible, making the method highly efficient and easy to deploy across pre-trained LALMs.

\section{EXPERIMENTS}

\subsection{Setup}

To systematically evaluate the effectiveness of our proposed MATA method, we conducted experiments on two mainstream audio reasoning benchmarks: MMAU\footnote{\href{https://github.com/Sakshi113/MMAU}{https://github.com/Sakshi113/MMAU}} and MMAR\footnote{\href{https://github.com/ddlBoJack/MMAR}{https://github.com/ddlBoJack/MMAR}}. For the MMAU dataset, since the initial version is no longer available, we used the Test-mini set with 1k audios of the updated version (v05.15.25).
We selected two state-of-the-art LALMs, Qwen2-Audio and Qwen2.5-Omni, as our primary baseline models. These models were chosen for their superior performance across various audio-textual multi-modal tasks, which provides a rigorous benchmark for evaluating the improvements offered by MATA. For hyperparameter settings, the enhancement strength \(\alpha\) was defaulted to 0.1, and the MATA intervention was applied to the LLM decoder layers from 10 to 20.

\subsection{Experimental results on the MMAU dataset}

Table \ref{MMAU} presents accuracy comparisons between our MATA method and mainstream baselines on MMAU (v05.15.25). The dataset categorizes audio inputs into three distinct tasks: sound, music, and speech. Given that the ground-truth answers for the full MMAU 9000 test set are not publicly available, all models were evaluated on the 1000 subset. As shown in Table 1, Qwen2.5-Omni achieves the best overall performance, with an average accuracy of 71.5\%. In contrast, Qwen2-Audio, once the top-performing open-source model in its original publication, exhibits a noticeable performance gap relative to the latest LALM, reflecting the rapid advancement of LALM architectures.

When integrated with our MATA method, both baseline models demonstrate significant performance improvements across all categories, which validates the effectiveness of our approach. For Qwen2-Audio (MATA), its accuracy increases from 59.4\% (reproduced) to 64.8\% on average. For the stronger baseline Qwen2.5-Omni, MATA still delivers consistent enhancements: its average accuracy rises from 71.1\% (reproduced) to 73.6\%, representing a 2.5\% improvement. Notably, the magnitude of improvement on Qwen2.5-Omni is slightly less pronounced compared to Qwen2-Audio, a phenomenon we attribute to its already robust baseline performance. Nonetheless, Qwen2.5-Omni (MATA) achieves state-of-the-art results in most categories, confirming MATA’s value even for cutting-edge models.

\begin{table}[t]
    \renewcommand{\arraystretch}{1}
    \setlength{\tabcolsep}{3pt} 
    \small
    \centering
    \vspace{-0.2cm}
    \begin{minipage}{0.48\textwidth}
    \centering
    \caption{Accuracies (\%) on MMAU (v05.15.25) benchmark. \textsuperscript{\textdagger} represents results reproduced in this work.}
    \begin{tabular}{@{}lcccc@{}}
        \hline
        \toprule
        \multirow{2}{*}{\textbf{Models}} & \multicolumn{4}{c}{\textbf{MMAU Test-mini}} \\
        \cmidrule(lr){2-5}
        & Sound & Music & Speech & Avg. \\
        \hline
        \multicolumn{5}{l}{\textit{\textbf{Baselines:}}} \\
        Gemini 2.0 Flash & 71.2 & 65.3 & \textbf{75.1} & 70.5 \\
        GPT-4o Audio & 64.6 & 56.3 & 66.7 & 62.5 \\
        LTU \cite{LTU} & 20.4 & 16.0 & 15.9 & 17.4 \\
        GAMA \cite{GAMA} & 31.8 & 17.7 & 12.9 & 20.8 \\
        SALMONN \cite{salmonn} & 41.1 & 37.1 & 26.4 & 34.9 \\
        Audio Flamingo 2 \cite{Audio-Flamingo2} & 71.5 & \textbf{71.0} & 44.7 & 62.4 \\
        Audio-Reasoner \cite{Audio-reasoner} & 67.9 & 69.2 & 66.1 & 67.7 \\
        Qwen2-Audio-7B-Instruct \cite{qwen2audio} & 67.3 & 56.3 & 55.3 & 59.6 \\
        Qwen2.5-Omni-7B \cite{qwen2.5omni} & \textbf{78.1} & 65.9 & 70.6 & \textbf{71.5} \\
        \hline
        \multicolumn{5}{l}{\textit{\textbf{Ours:}}} \\
        Qwen2-Audio-7B-Instruct\textsuperscript{\textdagger} & 67.1 & 55.8 & 55.4 & 59.4 \\
        \quad + MATA & 71.2 & 61.4 & 61.8 & 64.8 \\
        Qwen2.5-Omni-7B\textsuperscript{\textdagger} & 77.8 & 64.7 & 70.9 & 71.1 \\
        \quad + MATA & \textbf{79.9} & \textbf{68.3} & \textbf{72.7} & \textbf{73.6} \\
        \hline
    \end{tabular}
    \vspace{-0.1cm}
    \label{MMAU}
    \end{minipage}
\end{table}

\begin{table*}[t]
\renewcommand{\arraystretch}{1}
\small
\centering
\caption
{Accuracies (\%) on MMAR benchmark. \textsuperscript{\textdagger} represents results reproduced in this work. Ke-Omni-R-7B denotes a reinforcement learning fine-tuned version of Qwen-2.5-Omni-7B.}
\resizebox{0.98\linewidth}{!}{%
\begin{tabular}{lcccccccc}
\toprule
\multirow{2.5}{*}{\textbf{Model}} & \multicolumn{3}{c}{\textbf{Single Modality }} &  \multicolumn{4}{c}{\textbf{Mixed Modalities}} & \multirow{2.5}{*}{\textbf{Avg (\%)}} \\ 
\cmidrule{2-8}
& \textbf{Sound} & \textbf{Music} & \textbf{Speech} & \textbf{Sound-Music} & \textbf{Sound-Speech} & \textbf{Music-Speech} & \textbf{Sound-Music-Speech} & \\
\midrule
\multicolumn{9}{l}{\textit{\textbf{Baselines:}}} \\
Gemini 2.0 Flash & \textbf{61.2} & \textbf{51.0} & \textbf{72.1} & \textbf{81.8} & \textbf{72.5} & 65.9 & 70.8 & \textbf{65.6}   \\ 
GPT-4o Audio & 53.9 & \textbf{51.0} & 70.4 &63.6 & \textbf{72.5} & 62.2 & \textbf{75.0} & 63.5 \\
LTU \cite{LTU} & 19.4 & 19.9 & 14.0 & 18.2 & 24.8 & 22.0 & 16.7 & 19.2 \\
GAMA \cite{GAMA} & 29.1 & 24.3 & 27.9 & 27.3 & 24.8 & 28.1 & 20.8 & 26.5 \\
SALMONN \cite{salmonn} & 30.3 & 31.1 & 34.7 & 9.1 & 34.9 & 35.4 & 41.7 & 33.2 \\
Audio Flamingo 2 \cite{Audio-Flamingo2} & 24.9 & 17.5 & 20.8 & 18.2 & 26.6 & 23.2 & 8.3 & 21.9  \\
Audio-CoT \cite{Audio-cot} & 35.8 & 25.2 & 34.0 & 9.1 & 30.7 & 30.5 & 37.5 & 31.3 \\
Audio-Reasoner \cite{Audio-reasoner} & 43.6 & 33.5 & 33.0 & 45.5 & 42.7 & 31.7 & 25.0 & 36.8  \\
Qwen2-Audio-7B-Instruct \cite{qwen2audio} & 33.3 & 24.3 & 32.3 & 9.1 & 31.2 & 30.5 & 25.0 & 30.0  \\
Qwen-2.5-Omni \cite{qwen2.5omni} & 58.8 & 40.8 & 59.9 & 54.6 & 61.9 & \textbf{67.1} & 58.3 & 56.7 \\
\midrule
\multicolumn{9}{l}{\textit{\textbf{Ours:}}} \\
Qwen-2.5-Omni-7B\textsuperscript{\textdagger} \cite{qwen2.5omni} & 55.8 & 43.7 & 60.5 & 54.6 & 63.3 & 59.8 & 54.2 & 56.6 \\
\quad + MATA & 55.8 & 53.4 & 65.0 & 45.5 & 68.4 & 63.4 & 54.2 & 61.2 \\
Ke-Omni-R-7B\textsuperscript{\textdagger} \cite{qwen2.5omni} & 65.5 & \textbf{54.9} & 64.6 & 63.6 & 71.1 & 64.6 & \textbf{62.5} & 64.1 \\
\quad + MATA & \textbf{66.7} & 53.9 & \textbf{70.1} & \textbf{72.7} & \textbf{73.9} & \textbf{69.5} & \textbf{62.5} & \textbf{66.8} \\
\bottomrule
\end{tabular}%
}
\label{MMAR}
\end{table*}

\subsection{Experimental results on the MMAR dataset}

Table \ref{MMAR} presents accuracy comparisons on the MMAR benchmark, which comprises 1,000 test audio clips covering both single modality and mixed modality audio reasoning tasks. Among baseline models, proprietary closed-source systems, particularly Gemini 2.0 Flash and GPT-4o Audio, significantly outperform open-source alternatives. Specifically, Gemini 2.0 Flash achieves the highest average accuracy of 65.6\%, while the top open-source model, Qwen2.5-Omni, lags behind by nearly 9\%. This notable gap underscores the performance disparity between existing closed and open LALMs.

Our MATA method consistently enhances model performance. For Qwen2.5-Omni, MATA increases average accuracy from 56.6\% (reproduced) to 61.2\%, with improvements across most categories. A slight decrease is noted in the sound-music mixed modality task; however, this is statistically negligible as this category contains only 11 audio clips, corresponding to a single misclassified sample.

Notably, Ke-Omni-R-7B\footnote{\href{https://github.com/shuaijiang/Ke-Omni-R}{https://github.com/shuaijiang/Ke-Omni-R}}, a RL-fine-tuned variant of Qwen2.5-Omni, exhibits strong baseline performance with a 64.1\% average accuracy. Applying MATA to this RL-enhanced model yields further significant gains, pushing its average accuracy to 66.8\%, surpassing Gemini 2.0 Flash (65.6\%) and marking the first time that an open-source model has outperformed closed-source counterparts on MMAR benchmark. This outcome highlights MATA’s versatility and compatibility with existing fine-tuning strategies, as it can further enhance the performance of models already optimized through techniques like RL.

\begin{table}[t]
    \renewcommand{\arraystretch}{1.1}
    \small
    \centering
    \begin{minipage}{0.48\textwidth}
    \centering
    \vspace{-0.2cm}
    \caption{Ablation studies with different layers and $\alpha$. Accuracies (\%) on MMAU (v05.15.25).}
    \begin{tabular}{@{}cccccc@{}}
        \toprule 
        \multirow{2}{*}{$\alpha$} & \multirow{2}{*}{Layers} & \multicolumn{4}{c}{\textbf{MMAU Test-mini}} \\
        \cmidrule(lr){3-6}
        & & Sound & Music & Speech & Avg. \\
        \midrule 
        - & - & 77.8 & 64.7 & 70.9 & 71.1 \\
        \midrule 
        0.05 & 10-20 & 78.7 & 67.7 & 71.8 & 72.7 \\
        0.10 & 10-20 & 79.9 & \textbf{68.3} & \textbf{72.7} & \textbf{73.6} \\
        0.15 & 10-20 & 79.9 & 67.4 & 72.4 & 73.2 \\
        0.10 & 0-10 & 0.3 & 0.3 & 2.1 & 0.9 \\
        0.10 & 20-28 & \textbf{80.2} & 67.4 & 66.7 & 71.4 \\
        0.10 & 0-28 & 0.0 & 0.3 & 3.9 & 1.4 \\
        \bottomrule 
    \end{tabular}
    \label{Ablation}
    \end{minipage}
\end{table}

\subsection{Ablation studies}

We conducted ablation studies using the original Qwen2.5-Omni as the baseline to investigate the impact of enhancement strength $\alpha$ and intervention layer range. Table \ref{Ablation} summarizes the results on the MMAU benchmark. Our analysis shows that a moderate $\alpha$ value of 0.10 yields the highest average accuracy (73.6\%), with consistent improvements across all categories. In contrast, smaller or larger $\alpha$ values lead to either limited gains or slight performance degradation, suggesting that excessive audio attention amplification may disrupt the delicate balance of multi-modal information integration.

For the layer analysis, intervening on intermediate layers yields the best results, aligning with our earlier observation that these layers are the primary site of audio-text fusion where audio attention is inherently underweighted. In contrast, intervention in early layers (0-10) or late layers (20-28) is relatively poor. Early layer intervention leads to a catastrophic performance drop, likely due to distorted initial contextual encoding. Late layer intervention fails to surpass the baseline, as critical multi-modal fusion has already been completed. This validates the necessity of targeting specific intermediate layers to achieve optimal results.

\section{Conclusions}

In this paper, we investigate the critical problem of audio-textual attention imbalance in LALMs, identifying that they disproportionately allocate attention to text over audio, particularly in the crucial intermediate layers for multi-modal fusion. To mitigate this issue, we propose MATA, a training-free technique that dynamically amplifies attention to audio tokens within the self-attention mechanism. Experiments on the MMAU benchmark demonstrate that MATA improves the average accuracy of Qwen2-Audio and Qwen2.5-Omni by 5.4\% and 2.5\%, respectively. On the MMAR benchmark, MATA boosts the performance of the RL-fine-tuned Ke-Omni-R-7B model from 64.1\% to 66.8\%, marking the first time an open-source model surpasses the proprietary Gemini 2.0 Flash. For future work, we plan to extend the MATA method to a broader range of LALM architectures and explore its application in other multi-modal models to further improve their audio processing capabilities.

\vfill\pagebreak

\bibliographystyle{IEEEbib}
\bibliography{strings,refs}

\begin{thebibliography}{10}

\bibitem{SurveyLMRM}
Yunxin Li et~al.,
\newblock ``Perception, reason, think, and plan: A survey on large multimodal reasoning models,''
\newblock {\em arXiv preprint arXiv:2505.04921}, 2025.

\bibitem{deepseekR1}
Daya Guo et~al.,
\newblock ``Deepseek-r1: Incentivizing reasoning capability in llms via reinforcement learning,''
\newblock {\em arXiv preprint arXiv:2501.12948}, 2025.

\bibitem{COT1}
Jason Wei et~al.,
\newblock ``Chain-of-thought prompting elicits reasoning in large language models,''
\newblock 2022, vol.~35, pp. 24824--24837.

\bibitem{visualCOT}
Hao Shao et~al.,
\newblock ``Visual cot: Advancing multi-modal language models with a comprehensive dataset and benchmark for chain-of-thought reasoning,''
\newblock 2024, vol.~37, pp. 8612--8642.

\bibitem{Audio-reasoner}
Zhifei Xie, Mingbao Lin, Zihang Liu, Pengcheng Wu, Shuicheng Yan, and Chunyan Miao,
\newblock ``Audio-reasoner: Improving reasoning capability in large audio language models,''
\newblock {\em arXiv preprint arXiv:2503.02318}, 2025.

\bibitem{Audio-Flamingo2}
Sreyan Ghosh, Zhifeng Kong, Sonal Kumar, S~Sakshi, Jaehyeon Kim, Wei Ping, Rafael Valle, Dinesh Manocha, and Bryan Catanzaro,
\newblock ``Audio flamingo 2: An audio-language model with long-audio understanding and expert reasoning abilities,''
\newblock in {\em International Conference on Machine Learning}, 2025.

\bibitem{Audio-cot}
Ziyang Ma, Zhuo Chen, Yuping Wang, Eng~Siong Chng, and Xie Chen,
\newblock ``Audio-cot: Exploring chain-of-thought reasoning in large audio language model,''
\newblock {\em IEEE Automatic Speech Recognition and Understanding Workshop (ASRU)}, 2025.

\bibitem{LTU}
Yuan Gong, Hongyin Luo, Alexander~H Liu, Leonid Karlinsky, and James Glass,
\newblock ``Listen, think, and understand,''
\newblock in {\em International Conference on Learning Representations}, 2024.

\bibitem{GAMA}
Sreyan Ghosh et~al.,
\newblock ``Gama: A large audio-language model with advanced audio understanding and complex reasoning abilities,''
\newblock in {\em Proceedings of the 2024 Conference on Empirical Methods in Natural Language Processing}, 2024, pp. 6288--6313.

\bibitem{salmonn}
Changli Tang, Wenyi Yu, Guangzhi Sun, Xianzhao Chen, Tian Tan, Wei Li, Lu~Lu, Zejun MA, and Chao Zhang,
\newblock ``Salmonn: Towards generic hearing abilities for large language models,''
\newblock in {\em International Conference on Learning Representations}, 2024.

\bibitem{qwen2audio}
Yunfei Chu et~al.,
\newblock ``Qwen2-audio technical report,''
\newblock {\em arXiv preprint arXiv:2407.10759}, 2024.

\bibitem{imagebias}
Chengzhi Liu, Zhongxing Xu, Qingyue Wei, Juncheng Wu, James Zou, Xin~Eric Wang, Yuyin Zhou, and Sheng Liu,
\newblock ``More thinking, less seeing? assessing amplified hallucination in multimodal reasoning models,''
\newblock {\em arXiv preprint arXiv:2505.21523}, 2025.

\bibitem{PAI}
Shi Liu, Kecheng Zheng, and Wei Chen,
\newblock ``Paying more attention to image: A training-free method for alleviating hallucination in lvlms,''
\newblock in {\em European Conference on Computer Vision}, 2024, pp. 125--140.

\bibitem{hallucinationinfer}
Nick McKenna, Tianyi Li, Liang Cheng, Mohammad~Javad Hosseini, Mark Johnson, and Mark Steedman,
\newblock ``Sources of hallucination by large language models on inference tasks,''
\newblock in {\em The 2023 Conference on Empirical Methods in Natural Language Processing}.

\bibitem{hallucinationsurvey}
Lei Huang et~al.,
\newblock ``A survey on hallucination in large language models: Principles, taxonomy, challenges, and open questions,''
\newblock {\em ACM Transactions on Information Systems}, vol. 43, no. 2, pp. 1--55, 2025.

\bibitem{qwen2.5omni}
Jin Xu et~al.,
\newblock ``Qwen2.5-omni technical report,''
\newblock {\em arXiv preprint arXiv:2503.20215}, 2025.

\bibitem{hallucinationenhance}
Chaoya Jiang, Haiyang Xu, Mengfan Dong, Jiaxing Chen, Wei Ye, Ming Yan, Qinghao Ye, Ji~Zhang, Fei Huang, and Shikun Zhang,
\newblock ``Hallucination augmented contrastive learning for multimodal large language model,''
\newblock in {\em Proceedings of the IEEE/CVF Conference on Computer Vision and Pattern Recognition}, 2024, pp. 27036--27046.

\bibitem{MLLMsurvey}
Shukang Yin, Chaoyou Fu, Sirui Zhao, Ke~Li, Xing Sun, Tong Xu, and Enhong Chen,
\newblock ``A survey on multimodal large language models,''
\newblock {\em National Science Review}, vol. 11, no. 12, pp. nwae403, 2024.

\bibitem{mmau}
S~Sakshi et~al.,
\newblock ``Mmau: A massive multi-task audio understanding and reasoning benchmark,''
\newblock in {\em International Conference on Learning Representations}, 2025.

\bibitem{mmar}
Ziyang Ma et~al.,
\newblock ``Mmar: A challenging benchmark for deep reasoning in speech, audio, music, and their mix,''
\newblock {\em arXiv preprint arXiv:2505.13032}, 2025.

\bibitem{baichuanomni}
Yadong Li, Jun Liu, Tao Zhang, Song Chen, Tianpeng Li, Zehuan Li, Lijun Liu, Lingfeng Ming, Guosheng Dong, Da~Pan, et~al.,
\newblock ``Baichuan-omni-1.5 technical report,''
\newblock {\em arXiv preprint arXiv:2501.15368}, 2025.

\bibitem{hallucinationbias}
Hanchao Liu, Wenyuan Xue, Yifei Chen, Dapeng Chen, Xiutian Zhao, Ke~Wang, Liping Hou, Rongjun Li, and Wei Peng,
\newblock ``A survey on hallucination in large vision-language models,''
\newblock {\em arXiv preprint arXiv:2402.00253}, 2024.

\bibitem{retrain_align}
Zhiqing Sun et~al.,
\newblock ``Aligning large multimodal models with factually augmented rlhf,''
\newblock in {\em Annual Meeting of the Association for Computational Linguistics}, 2024.

\bibitem{omnir1}
Andrew Rouditchenko, Saurabhchand Bhati, Edson Araujo, Samuel Thomas, Hilde Kuehne, Rogerio Feris, and James Glass,
\newblock ``Omni-r1: Do you really need audio to fine-tune your audio llm?,''
\newblock {\em arXiv preprint arXiv:2505.09439}, 2025.

\end{thebibliography}

\end{document}